\documentclass[prl, twocolumn, tightenlines, twoside, secnumarabic, superscriptaddress, preprintnumbers, nofootinbib, notitlepage]{revtex4-1}
\usepackage{graphicx}
\usepackage{epstopdf}
\usepackage{amsmath}
\usepackage{amsfonts}
\usepackage{amssymb}
\usepackage{appendix}
\usepackage{comment}
\usepackage{bbold}
\usepackage{color}
\usepackage{slashed}
\usepackage{setspace}
\usepackage{footnote}
\usepackage{multirow}
\usepackage{mathrsfs}
\usepackage{feynmp-auto}
\usepackage{hyperref}
\usepackage{cleveref}
\usepackage{comment}

\usepackage{wrapfig}

\definecolor{niceblue}{rgb}{0.388235, 0.627451, 0.847059}
\definecolor{nicered}{rgb}{0.7,0.1,0.1}
\definecolor{nicegreen}{rgb}{0.1,0.5,0.1}
\definecolor{blue-green}{rgb}{0.0, 0.87, 0.87}
\hypersetup{colorlinks,citecolor= nicegreen,linkcolor= nicered}

\DeclareMathOperator{\Tr}{Tr}

\newcommand{\trm}[1]{\textrm{#1}}

\newcommand{\Mpc}{\trm{\Mpc}}
\newcommand{\yr}{\trm{\yr}}
\newcommand{\eV}{\trm{\eV}}

\newcommand{\nuh}{{N}}
\newcommand{\nul}{{\nu_L}}

\newcommand{\ee}{{e^+e^-}}
\newcommand{\bq}{\mathbf{q}}
\newcommand{\bk}{\mathbf{k}}
\newcommand{\bl}{\mathbf{l}}
\renewcommand{\slash}[2]{#1\hskip #2 em/}
\newcommand{\kslash}{\slash{k}{-0.5}}
\newcommand{\lslash}{\slash{l}{-0.38}}
\newcommand{\pslash}{\slash{p}{-0.5}}
\newcommand{\sslash}{\slash{s}{-0.5}}
\newcommand{\qslash}{\slash{q}{-0.5}}

\begin{document}

\singlespacing
\allowdisplaybreaks

\title{Heavy Neutral Lepton Decay Searches using Solar Neutrinos}

\author{Patrick Huber}
\author{Yulun Li}
\affiliation{Center for Neutrino Physics, Department of Physics, Virginia Tech, Blacksburg, VA 24061, USA}


\begin{abstract}
We study the sensitivity to the decay of a heavy neutral lepton into $\ee$-pairs using the solar boron-8 neutrino flux as source. We provide a fully differential cross section for this  process including the interference of neutral and charged current amplitudes. We revisit a previous bound from Borexino and make predicitions for the expected sensitivity in future large liquid noble gas detectors, like XLZD, Argo and DUNE, as well as high-resolution scintillator detectors based on the LiquidO technology. We find that more than two orders of magnitude improvement in mixing angle reach is possible relative to existing bounds.
\end{abstract}

\maketitle

\section{Introduction}

The Standard Model contains three neutral leptons, neutrinos, which we now know to be massive. The invisible width of the $Z$-boson excludes additional neutrinos with a mass below $m_Z/2$ coupling to it with the usual weak strength. However, neutral fermions that are SM singlets, {\it i.e.} do not couple to the $Z$, are possible over a wide range of masses. These hypothetical SM singlet states are usually called either sterile neutrinos or neutral heavy leptons (NHL). The absence of a coupling to the $Z$ can be either due to a lack of the corresponding coupling altogether or due to the wrong helicity. Apart from not being excluded by data, theory motivations for the existence of sterile neutrinos/HNLs can be found ranging from neutrino mass generation, over various cosmological anomalies that get ameliorated to leptogenesis. These new states can mix with the usual SM neutrinos, and this mixing provides the means to test their existence in a model-independent way.

If at least one SM neutrino mass states $i$ had sufficient mass $m_i>2m_e$, then the following decay would be possible $\nu_i\rightarrow\nu_j + e^+ +e^-$ and would receive both a neutral current (NC) and charged current (CC) contribution, which interfere. For HNLs this decay is entirely described by the SM and only the mixing with the SM neutrinos and the HNL mass are new parameters; we will derive limits in these two parameters.

In this paper we consider HNLs with $m>2m_e\simeq 1\,$MeV and consider their decay to the final state described above containing a $\ee$-pair. The neutrino source we consider is the solar flux of boron-8 neutrinos, which extends up about 16\,MeV. We first will compute the differential cross section for this final state, and present a result which correctly accounts for the NC/CC interference.  This final state is quite different from the elastic scattering Standard Model process $\nu_e + e\rightarrow \nu_e + e$ boron-8 neutrinos also will undergo. In a detector like Borexino (or DUNE) these elastic scattering events present an irreducible background. The signature of the $\ee$ final state will be the sum energy of the $\ee$-pair and the annihilation gammas resulting in a distortion of the energy spectrum. We call this type of analysis calorimetric. We revisit the previous Borexino limit~\cite{Borexino:2013bot} using the newly derived cross section and evaluate the potential sensivity of DUNE. Here we assume the detectors can be improved to be sensitive to boron-8 solar neutrinos~\cite{Capozzi:2018dat}.

However, in a detector with sufficient spatial resolution and particle identification this final state presents a very clean signal: the $e^-$ and $e^+$ will lead to two tracks and the $e^+$ furthermore will result in two back-to-back 0.511\,MeV gammas. We will show that the opening angle between the $\ee$ is sufficiently large, see the appendix, throughout the parameter space to allow for this signature being visible. The only natural background then is pair production from high-energy gammas. We call this type of analysis tracking-based. Future large liquid noble gas detectors made from xenon, for example XLZD~\cite{Aalbers:2022dzr} or argon, for example Argo~\cite{McDonald:2024osu}, are obvious candidates for this detection mode  as is the newly proposed LiquidO scintillation technology~\cite{LiquidO:2019mxd}. We also will derive the expected sensitivities using the tracking-based detection  mode.

\section{SM decays of HNL}

We note  that there are other available decay modes for an MeV-scale HNL aside from $\nu e^+ e^-$, see Ref.~\cite{Gorbunov:2007ak}:
\begin{eqnarray}
	& \Gamma(N \to \sum_{\alpha,\beta} \nu_\alpha\nu_\beta\overline{\nu}_\beta) = \dfrac{G_F^2 m_N^5}{192 \pi^3} \sum_\alpha |U_{\alpha}|^2  & \\
	& \Gamma(N \to \sum_\alpha \nu_\alpha \gamma) = \dfrac{9 \alpha_{\text{EM}}G_F^2 m_N^5}{512 \pi^4} \sum_\alpha |U_{\alpha}|^2 &
\end{eqnarray}
The invisible mode ($N\rightarrow \nu\nu\overline{\nu}$) is not included in our analysis since it leaves no experimentally accessible signature. Also, we did not include the radiative decay mode ($N\rightarrow\nu\gamma$) in our calculation because the radiative decay time is much longer than the flight time from the core of the Sun to Earth. For example, for $m_N=5$ MeV and $|U_{eH}|^2=1$, the decay time is $~10^{10}$ s. Consequently, the mass parameter space we are interested in has a lower limit of $2m_e$. 

We calculated the closed-form differential decay width for our analysis, where we sum over the spins of the final states and average over the two possible states of $N$. We integrate the matrix element over the momenta over the outgoing electron and positron. Employing the usual kinematics, we obtain the differential decay width in terms of $Q$, the momentum transfer to $\ee$-pair and $\cos\theta$, the angle of emission of $\nu_e$ relative to $N$'s polarization direction.  Putting everything together, we have:
\begin{widetext}

\begin{eqnarray}
    \frac{d^2\overline{\Gamma}}{d l^0 d\cos\theta}
    &=&2(1-Q^2)^2\sqrt{1-\frac{4m_e^2}{Q^2}}\frac{1}{Q^2}\bigg\{\cr
    &&\qquad\qquad\bigg[X\underbrace{\bigg(Q^2+2Q^4-2m_e^2(Q^2-1)\bigg)}_{\displaystyle \mathrm{CC}\;\mathrm{Contribution}}-6Z Q^2 m_e^2\bigg]\cr
    &&-|\Vec{s}|\cos\theta\bigg[X\underbrace{\bigg(Q^2-2Q^4+2m_e^2(1+Q^2)\bigg)}_{\displaystyle \mathrm{CC}\;\mathrm{Contribution}}+6Z Q^2 m_e^2\bigg]\bigg\}
\end{eqnarray}

\end{widetext}
where $X=[(g_V+1)^2+(g_A+1)^2]$ and $Z=[(g_V+1)^2-(g_A+1)^2]$. The term that is multiplied by $X$ is from the charged-current contribution which agrees with the previous result in Ref.~\cite{Shrock:1981wq}. The coefficients $X$ and $Z$ are the result of interference and we believe have not been reported before in the literature, the detailed expression can be found in the appendix.


\subsection{Integrated Decay Rates}

We would like to determine the rate at which solar HNLs decay in a detector on Earth or equivalently, the number of decays that occur in a specified volume during a specified time. The number of decays  is given as follows:
\begin{equation}
	\label{eq:total_decays}
	N_{\rm tot}= VT \times \int d E_{N} \, \frac{d\widetilde{\Phi}_\odot(E_{N})}{dE_{N}} \frac{m_{N} }{p_{N}}\frac{\Gamma^{e}|U_{eN}|^2}{\Gamma_{\rm tot}} e^{-\frac{D_\odot m_{N} \Gamma_{\rm tot}}{E_{N}\beta c}},
\end{equation}
where we define the following quantities: $VT$ is the total exposure, {\it i.e.} the product of the detector volume and the duration of the observation; $m_{N}$ is the HNL mass, $E_{N}$ is its (lab-frame) energy and $p_{N}$ the corresponding momentum; $\Gamma_{\rm tot}$ is the total (rest-frame) width of the HNL, $\Gamma^{e}$ is the width of the $N\rightarrow \nu e^+ e^-$, and $\frac{\Gamma^{e}}{\Gamma_{\rm tot}}$ is the branching ratio. Now the  \emph{inverse decay length} of the HNLs is given by $ \Gamma_{\rm tot}/(\gamma\, v_N)=(m_N/E_N)(\Gamma_{\rm tot}/\beta c)$, and the fraction of HNLs that survive the journey from the Sun to Earth is $e^{-\frac{D_\odot m_{N} \Gamma_{\rm tot}}{E_{N}\beta c}}$, with $D_\odot = 1.5 \times 10^{13}$ cm the mean Earth-Sun distance. This yields a flux per unit area per unit time of HNLs
	\begin{equation}
		\frac{d\widetilde{\Phi}_\odot}{dE_{N}} = |U_{eN}|^2 \frac{p_{N}}{E_{N}} \frac{d\Phi_\odot}{dE_\nu},
	\end{equation}
where $d\Phi_\odot/dE_\nu$ is the standard solar neutrino flux at Earth (which does not incorporate solar flavor conversion). Note, the lower bound on the integral in Eq.~\eqref{eq:total_decays} is given by $m_{N}$.

\subsection{Differential Decay Rates}

To obtain the experimental signature we need to determine the \emph{differential decay rate} in the laboratory frame for this process. To be able to articulate this, we make note of the following relationships among the lab-frame and rest-frame observables. The lab-frame neutrino energy, $E_\nul$, is related to rest frame neutrino energy $\epsilon$ via 
$$ E_\nul=\gamma(\epsilon+\beta\cos\theta\epsilon)=\frac{E_\nuh}{m_\nuh}\epsilon(1+\beta\cos\theta).$$ The maximum outgoing neutrino energy in the
rest frame is
	\begin{eqnarray}
    Q^2&=&m_\nuh^2-2m_\nuh \epsilon_{\mathrm{max}}\overset{!}{=}(2m_e)^2\cr
    \epsilon_{\mathrm{max}}&=&\frac{m_\nuh^2-4m_e^2}{2m_\nuh}.
\end{eqnarray}
We further find that $\cos\theta$ is inversely proportional to $\epsilon$, thus for $\epsilon$ to have a maximum, $\cos\theta$ needs to have a minimum
 \begin{equation}
    (\cos\theta)_{\mathrm{min}}=\frac{1}{\beta}\bigg(\frac{m_\nuh}{\epsilon_{\mathrm{max}}}(1-\frac{E_{e^+}+E_{e^-}}{E_\nuh})-1\bigg).
\end{equation}
To calculate the differential decay rate, we need to transform from $\frac{dN}{d E_\nul}$ to $\frac{dN}{d E}$ where $E$ is the energy of the $\ee$-pair. 
\begin{eqnarray}
N_\ee\cr&=&N_\nul=\int_{\cos\theta_{\mathrm{min}}}^1 d\cos\theta \int d\epsilon \frac{d N_\nul}{d\epsilon d\cos\theta}\cr
&=&\int_{\cos\theta_{\mathrm{min}}}^1 d\cos\theta \int d E \frac{m_\nuh}{E_\nuh(1+\beta\cos\theta)}\frac{d N_\nul}{d\epsilon d\cos\theta}\cr
\Rightarrow &&\frac{d N_\ee}{d E}\cr&=&\int_{\cos\theta_{\mathrm{min}}}^1 d\cos\theta \frac{m_\nuh}{E_\nuh(1+\beta\cos\theta)}\frac{d N_\nul}{d\epsilon d\cos\theta}.
\end{eqnarray}

In addition, we multiply the time $\frac{1}{\beta c}$ that $\nuh$ spends in the detector per unit length, along with other unit conversions, such that the spectra describe the $\ee$ that are produced inside the detector. With this in hand, we write the visible energy spectrum arising from the decays of solar HNLs via $N \to \nu_\alpha e^+ e^-$ as follows:
\begin{widetext}
\begin{equation}
    \frac{d N_\ee}{d E}=\int d E_\nuh \int_{\cos\theta_\mathrm{min}}^1 d\cos\theta \frac{\Gamma^e}{\Gamma^\mathrm{tot}}\frac{d\widetilde{\Phi}_\odot}{dE_N} e^{-\frac{D_\odot m_{N} \Gamma_{\rm tot}}{p_{N}}}\frac{m_\nuh}{E_\nuh(1+\beta\cos\theta)}\frac{d N_\nul}{d\epsilon d\cos\theta}|U_{e H}|^2\frac{1}{\beta c}
\end{equation}
\end{widetext}

\section{Existing Constraints on Heavy Neutral Leptons}

\begin{figure}
\includegraphics[width=\columnwidth]{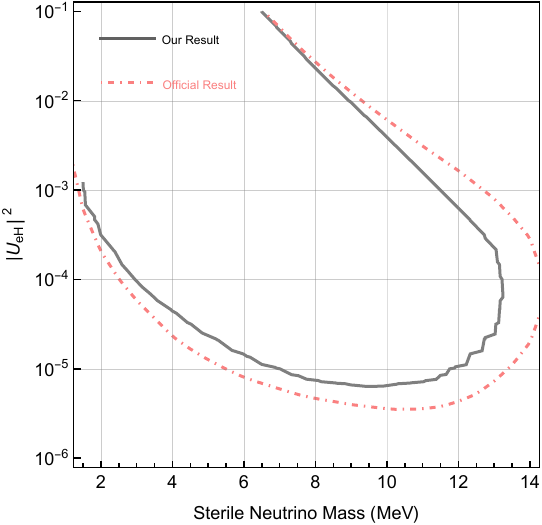}
\caption{The results of our reanalysis of Borexino data \cite{Borexino:2013bot}. Our 90\% confidence exclusion is shown in a dark grey solid line. The light red dot-dashed curve corresponds to the 90\% confidence limit published in Ref.~\cite{Borexino:2013bot}.}
    \label{fig:our_bxno_result}
\end{figure}

A search for $\ee$-decay mode has been previously conducted by the Borexino collaboration~\cite{Borexino:2013bot} using solar boron-8 neutrinos with an exposure of 0.122\,kt\,yr. In the Borexino detector a $\ee$-pair cannot be resolved into its constituents or otherwise distinguished at the event-by-event level from any other interaction in the detector. Therefore, the $\nu_e+e^-\rightarrow\nu_e+e^-$ elastic scattering events from the boron-8 flux present an irreducible background for this search, {\it i.e.} this is a calorimetric analysis. We take the electron recoil data from Fig.~4 of Ref.~\cite{Borexino:2013bot} and perform a binned likelihood fit for signal with this data as background. In Fig~\ref{fig:our_bxno_result} we compare our result for the \emph{same} data set. Due to the interference of the NC and CC contributions, our limit is somewhat weaker than the Borexino one, highlighting that the interference of the neutral and charged current amplitudes is non-negligible.

There are many terrestrial experiments sensitive to heavy neutral leptons as well. Here, we briefly summarize the most relevant constraints on MeV-scale heavy neutral leptons. For example, there are results of a combined analysis of superallowed decays from Ref.~\cite{Deutsch:1990ut}. 
In the mass range of 1-10\,Mev the strongest constraints come from Bugey \cite{Hagner:1995bn}, being the dominant constraint from nuclear reactors. Previous constraints from reactor experiments at G\"osgen \cite{Oberauer:1987mg} and Rovno \cite{Kopeikin:1990vp}  have been superseded by this constraint from Bugey. The pion beta-decay experiment PIENU \cite{Bryman:2019bjg} is sensitive to the ratio BR$(\pi^+\to e^+\nu_e)$/BR$(\pi^+\to\mu^+\nu_\mu)$ and can find bumps in $\pi_{e2}$ decays. A HNL can imprint itself on muon decay via modifications to the electron spectrum, {\it i.e.}, by modifying the Michel parameters. In Ref.~\cite{deGouvea:2015euy} the authors studied the impact of an HNL on observations from the TWIST experiment \cite{TWIST:2009zko, Bayes:2013esa}; their constraint is subdominant to nuclear decays and to PIENU. We will not address the issue of HNLs as dark matter, see Refs.~\cite{Drewes:2016upu, Boyarsky:2018tvu}.  Moreover, we acknowledge powerful constraints on the existence of HNLs from Big Bang Nucleosynthesis (BBN) \cite{Sabti:2020yrt} covering the parameter space considered here. However, these constraints are sensitive to difficult-to-constrain inputs like the neutrino asymmetry priory to BBN and thus are complimentary to the results presented here.

\begin{figure}[!t]
    \centering
        \includegraphics[width=\linewidth]{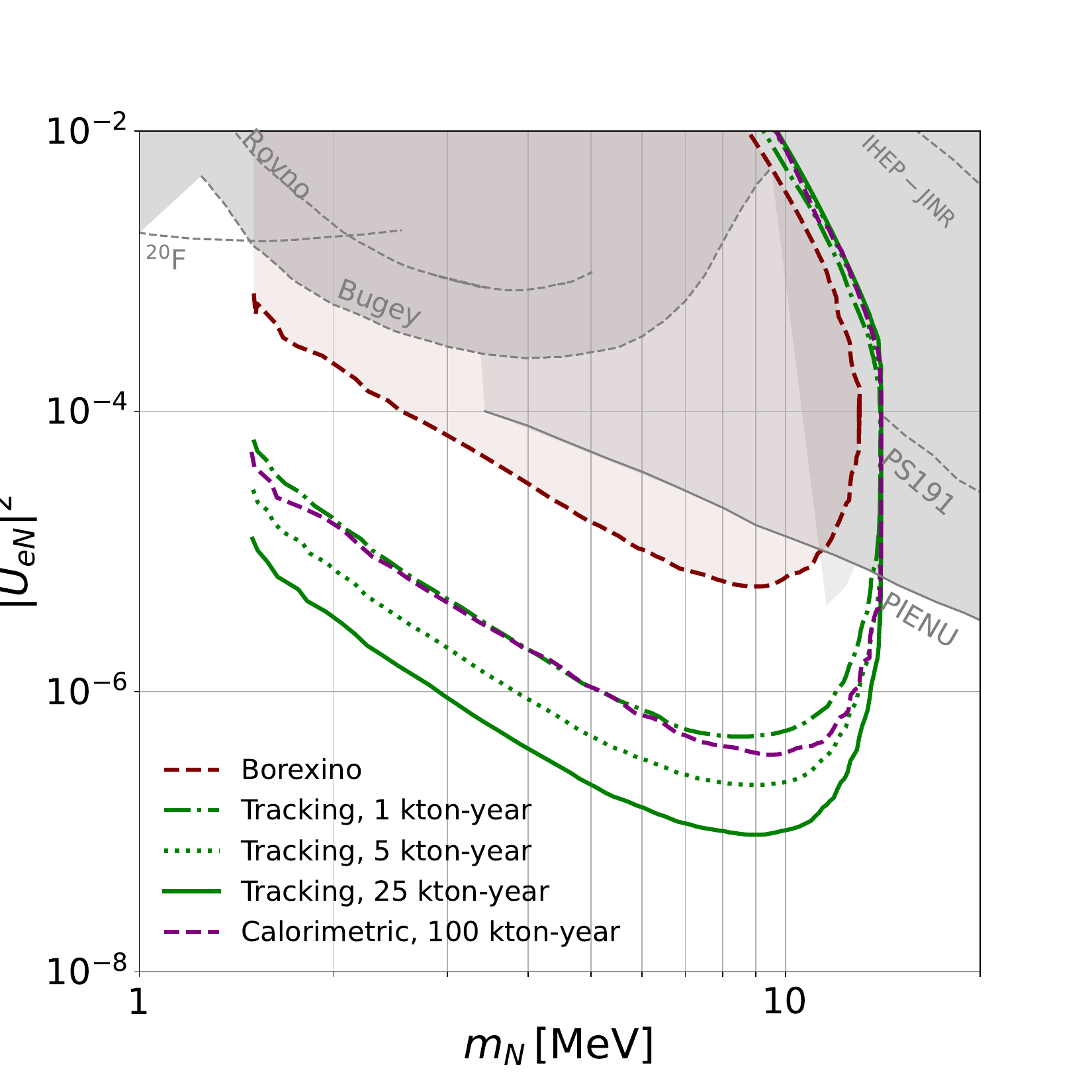}
    \caption{Exclusions on and sensitivities to the HNL parameter space \cite{Baranov:1992vq,Bernardi:1987ek,Calaprice:1983qn,Derbin:1993wy,Hagner:1995bn,PIENU:2017wbj,Bolton:2019pcu}. All constraints are at 90\% C.L. The gray regions correspond to various existing constraints; see text for more details. 
    }
    \label{fig:money_plot}
\end{figure}

\section{Future Sensitivities}

Improvements on the Borexino bound can happen for two reasons, either an increase in statistics by having a larger detector and/or by having detectors that can tag $\ee$-pairs and thus can, on an event-by-event basis, reject the solar neutrino elastic electron scattering background.

In terms of increasing statistics, size matters and the largest currently under construction liquid argon detectors are the ones of the DUNE experiment. The DUNE detectors in their default configuration are not sensitive to relatively rare signals at the 10\,MeV level. However, in  Ref.~\cite{Capozzi:2018dat} the case was made that there is a strong physics case to extend the sensitivity to lower energies and a concept was outlined how to do this.  We therefore will evaluate the sensitivity of a low-energy improved DUNE detector. Our analysis of solar HNL production at DUNE is similar to that of Borexino data. The expected solar $^8\mathrm{B}$ elastic scattering (ES) signal $\nu_{e,\mu,\tau}+e^-$ is digitized from Ref.~\cite{Capozzi:2018dat}. For the expected HNL production, we assume that there is no additional background. We further assume 100 kton-year exposure for the liquid argon detector and above 5 MeV, we assumed $7\%$ energy-independent energy resolution.  The resulting $90\%$ C.L. sensitivity is shown as purple, solid line in Fig.~\ref{fig:money_plot}

Detectors that can tag $\ee$-pairs need to have a high spatial resolution and a low enough energy threshold to identify the 2 annihilation gammas with only 0.511\,MeV, each, where the first Compton-scatter events usually have only fraction of this already low energy. One class of detectors that can achieve that, are liquid noble gas dark matter detectors and the third generation of these experiment can reach masses comparable to or exceeding that of Borexino, {\it e.g} XLZD~\cite{Aalbers:2022dzr} with 50\,tons of xenon and Argo~\cite{McDonald:2024osu} with 400\,tons of argon. These detectors offer excellent sensitivity to the creation of a $e^+e^-$-pair as they can resolve the resulting tracks originating from a common vertex. They also usually feature keV-level thresholds and hence can identify annihilation gammas, thus they can reliably tag $\ee$ pairs with high efficiency. The only background resulting in $\ee$-pairs is pair conversion of high-energy gammas from either natural radioactivity or cosmogenic events. The noble gases used in  these detectors are practically free from radioactive contaminants and hence high-energy gammas have to enter from the outside. Xenon has a very high stopping power for gammas due to the high atomic number and density and thus, xenon-based detectors have excellent self-shielding. For argon the self-shielding is less pronounced due to a lower density and smaller atomic number, but due to the better availability of argon compared to xenon, argon detectors tend to be much larger, compensating for the lower stopping power. Argon-39 a radioactive, cosmogenic isotope of argon beta-decays with an energy too low to ever lead to pair-production. Moreover, for any dark matter experiment ancient underground argon with a very low concentration of argon-39 will be used. In principle, argon-39 beta decay could interfere with the identification of the annihilation gammas but a detailed detector study is required to address this question. We assume that close spatial/temporal coincidence cuts around the two tracks from the $\ee$-pair can practically solve this problem.  Under these assumptions the 90\% C.L. sensitivity corresponds to the detection of 2.3 events. We present limits for 1 and 5 kton-yr  total exposures in dashed green, in Fig.~\ref{fig:money_plot} corresponding to 10--20 years of running time.

Finally, LiquidO~\cite{LiquidO:2019mxd} is a relatively new detector concept based on stochastic light confinement: in a scintillating medium where the scattering length is short compared to the absorption length, scintillation light gets trapped around its origin. The trapped light can be read out with wavelength-shifting optical fibers with a spatial resolution determined by the spacing of the optical fibers. Recently, prototypes of this technology provided a proof-of-concept~\cite{LiquidO:2024piw,LiquidO:2025qia}. This technology can conceivably be scaled in the kiloton range. We know little about the performance of kiloton-sized detectors of this type, but positron-tagging and tagging of the relatively complex sequence involved in neutrino capture on indium-115 is advertised~\cite{LiquidO:2019mxd}. We therefore, to illustrate the potential of this technology assume that background-free tagging of $\ee$-pairs in kiloton-scale detectors is possible and show a sensitivity corresponding to a 25 kton-yr exposure in  Fig.~\ref{fig:money_plot}.


\section{Summary \& Outlook}

Heavy neutral leptons are a common ingredient in many models of new physics and we revisited searches for them in the mass range where the SM decay into $\ee$ is inevitable. We provided a complete expression for the fully differential decay rate including the neutral/charge current interference. We then use this expression to compute the rates expected in Borexino from the solar boron-8 neutrino flux and compare our result to a previous analysis by the collaboration~\cite{Borexino:2013bot}. We find indeed that the inference term changes the limit somewhat. We then proceed to evaluate the sensitivity of future liquid noble gas detectors, either using xenon or argon, and compare the expected sensitivity to existing bounds. We find that a low-energy improved DUNE allows for a calorimetric analysis similar to what was done for Borexino and would result in a limit about 30 times more stringent. Dark matter detectors like XLZD and Argo are much smaller than DUNE but provide much better spatial and energy resolution and hence can perform a tracking analysis. These experiments could improve the mixing angle sensitivity by about a factor of 30--80 depending on the exposure. In the case where very large exposures become possible, for instance by using LiquidO technology, even better sensitivities can be achieved, improving limits by more than  two orders of magnitude.  Any of these detectors will be built for physics cases independent of heavy neutral leptons and the Sun will continue to provide boron-8 neutrinos, therefore, these searches require minimal resources once the detectors are built.


\section*{Acknowledgments}

We thank  J.M.~Berryman for collaboration during the early stages of this work. This work was supported by the U.S. Department of Energy Office of Science awards  DE-SC00018327 and DE-SC00020262.

\bibliography{HNL_bib}{}

\begin{thebibliography}{26}%
\makeatletter
\providecommand \@ifxundefined [1]{%
 \@ifx{#1\undefined}
}%
\providecommand \@ifnum [1]{%
 \ifnum #1\expandafter \@firstoftwo
 \else \expandafter \@secondoftwo
 \fi
}%
\providecommand \@ifx [1]{%
 \ifx #1\expandafter \@firstoftwo
 \else \expandafter \@secondoftwo
 \fi
}%
\providecommand \natexlab [1]{#1}%
\providecommand \enquote  [1]{``#1''}%
\providecommand \bibnamefont  [1]{#1}%
\providecommand \bibfnamefont [1]{#1}%
\providecommand \citenamefont [1]{#1}%
\providecommand \href@noop [0]{\@secondoftwo}%
\providecommand \href [0]{\begingroup \@sanitize@url \@href}%
\providecommand \@href[1]{\@@startlink{#1}\@@href}%
\providecommand \@@href[1]{\endgroup#1\@@endlink}%
\providecommand \@sanitize@url [0]{\catcode `\\12\catcode `\$12\catcode `\&12\catcode `\#12\catcode `\^12\catcode `\_12\catcode `\%12\relax}%
\providecommand \@@startlink[1]{}%
\providecommand \@@endlink[0]{}%
\providecommand \url  [0]{\begingroup\@sanitize@url \@url }%
\providecommand \@url [1]{\endgroup\@href {#1}{\urlprefix }}%
\providecommand \urlprefix  [0]{URL }%
\providecommand \Eprint [0]{\href }%
\providecommand \doibase [0]{http://dx.doi.org/}%
\providecommand \selectlanguage [0]{\@gobble}%
\providecommand \bibinfo  [0]{\@secondoftwo}%
\providecommand \bibfield  [0]{\@secondoftwo}%
\providecommand \translation [1]{[#1]}%
\providecommand \BibitemOpen [0]{}%
\providecommand \bibitemStop [0]{}%
\providecommand \bibitemNoStop [0]{.\EOS\space}%
\providecommand \EOS [0]{\spacefactor3000\relax}%
\providecommand \BibitemShut  [1]{\csname bibitem#1\endcsname}%
\let\auto@bib@innerbib\@empty
\bibitem [{\citenamefont {Bellini}\ \emph {et~al.}(2013)\citenamefont {Bellini} \emph {et~al.}}]{Borexino:2013bot}%
  \BibitemOpen
  \bibfield  {author} {\bibinfo {author} {\bibfnamefont {G.}~\bibnamefont {Bellini}} \emph {et~al.} (\bibinfo {collaboration} {Borexino}),\ }\href {\doibase 10.1103/PhysRevD.88.072010} {\bibfield  {journal} {\bibinfo  {journal} {Phys. Rev. D}\ }\textbf {\bibinfo {volume} {88}},\ \bibinfo {pages} {072010} (\bibinfo {year} {2013})},\ \Eprint {http://arxiv.org/abs/1311.5347} {arXiv:1311.5347 [hep-ex]} \BibitemShut {NoStop}%
\bibitem [{\citenamefont {Capozzi}\ \emph {et~al.}(2019)\citenamefont {Capozzi}, \citenamefont {Li}, \citenamefont {Zhu},\ and\ \citenamefont {Beacom}}]{Capozzi:2018dat}%
  \BibitemOpen
  \bibfield  {author} {\bibinfo {author} {\bibfnamefont {F.}~\bibnamefont {Capozzi}}, \bibinfo {author} {\bibfnamefont {S.~W.}\ \bibnamefont {Li}}, \bibinfo {author} {\bibfnamefont {G.}~\bibnamefont {Zhu}}, \ and\ \bibinfo {author} {\bibfnamefont {J.~F.}\ \bibnamefont {Beacom}},\ }\href {\doibase 10.1103/PhysRevLett.123.131803} {\bibfield  {journal} {\bibinfo  {journal} {Phys. Rev. Lett.}\ }\textbf {\bibinfo {volume} {123}},\ \bibinfo {pages} {131803} (\bibinfo {year} {2019})},\ \Eprint {http://arxiv.org/abs/1808.08232} {arXiv:1808.08232 [hep-ph]} \BibitemShut {NoStop}%
\bibitem [{\citenamefont {Aalbers}\ \emph {et~al.}(2023)\citenamefont {Aalbers} \emph {et~al.}}]{Aalbers:2022dzr}%
  \BibitemOpen
  \bibfield  {author} {\bibinfo {author} {\bibfnamefont {J.}~\bibnamefont {Aalbers}} \emph {et~al.},\ }\href {\doibase 10.1088/1361-6471/ac841a} {\bibfield  {journal} {\bibinfo  {journal} {J. Phys. G}\ }\textbf {\bibinfo {volume} {50}},\ \bibinfo {pages} {013001} (\bibinfo {year} {2023})},\ \Eprint {http://arxiv.org/abs/2203.02309} {arXiv:2203.02309 [physics.ins-det]} \BibitemShut {NoStop}%
\bibitem [{\citenamefont {McDonald}(2024)}]{McDonald:2024osu}%
  \BibitemOpen
  \bibfield  {author} {\bibinfo {author} {\bibfnamefont {A.~B.}\ \bibnamefont {McDonald}},\ }\href {\doibase 10.1016/j.nuclphysb.2024.116436} {\bibfield  {journal} {\bibinfo  {journal} {Nucl. Phys. B}\ }\textbf {\bibinfo {volume} {1003}},\ \bibinfo {pages} {116436} (\bibinfo {year} {2024})}\BibitemShut {NoStop}%
\bibitem [{\citenamefont {Cabrera}\ \emph {et~al.}(2021)\citenamefont {Cabrera} \emph {et~al.}}]{LiquidO:2019mxd}%
  \BibitemOpen
  \bibfield  {author} {\bibinfo {author} {\bibfnamefont {A.}~\bibnamefont {Cabrera}} \emph {et~al.} (\bibinfo {collaboration} {LiquidO}),\ }\href {\doibase 10.1038/s42005-021-00763-5} {\bibfield  {journal} {\bibinfo  {journal} {Commun. Phys.}\ }\textbf {\bibinfo {volume} {4}},\ \bibinfo {pages} {273} (\bibinfo {year} {2021})},\ \Eprint {http://arxiv.org/abs/1908.02859} {arXiv:1908.02859 [physics.ins-det]} \BibitemShut {NoStop}%
\bibitem [{\citenamefont {Gorbunov}\ and\ \citenamefont {Shaposhnikov}(2007)}]{Gorbunov:2007ak}%
  \BibitemOpen
  \bibfield  {author} {\bibinfo {author} {\bibfnamefont {D.}~\bibnamefont {Gorbunov}}\ and\ \bibinfo {author} {\bibfnamefont {M.}~\bibnamefont {Shaposhnikov}},\ }\href {\doibase 10.1088/1126-6708/2007/10/015} {\bibfield  {journal} {\bibinfo  {journal} {JHEP}\ }\textbf {\bibinfo {volume} {10}},\ \bibinfo {pages} {015} (\bibinfo {year} {2007})},\ \bibinfo {note} {[Erratum: JHEP 11, 101 (2013)]},\ \Eprint {http://arxiv.org/abs/0705.1729} {arXiv:0705.1729 [hep-ph]} \BibitemShut {NoStop}%
\bibitem [{\citenamefont {Shrock}(1981)}]{Shrock:1981wq}%
  \BibitemOpen
  \bibfield  {author} {\bibinfo {author} {\bibfnamefont {R.~E.}\ \bibnamefont {Shrock}},\ }\href {\doibase 10.1103/PhysRevD.24.1275} {\bibfield  {journal} {\bibinfo  {journal} {Phys. Rev. D}\ }\textbf {\bibinfo {volume} {24}},\ \bibinfo {pages} {1275} (\bibinfo {year} {1981})}\BibitemShut {NoStop}%
\bibitem [{\citenamefont {Deutsch}\ \emph {et~al.}(1990)\citenamefont {Deutsch}, \citenamefont {Lebrun},\ and\ \citenamefont {Prieels}}]{Deutsch:1990ut}%
  \BibitemOpen
  \bibfield  {author} {\bibinfo {author} {\bibfnamefont {J.}~\bibnamefont {Deutsch}}, \bibinfo {author} {\bibfnamefont {M.}~\bibnamefont {Lebrun}}, \ and\ \bibinfo {author} {\bibfnamefont {R.}~\bibnamefont {Prieels}},\ }\href {\doibase 10.1016/0375-9474(90)90541-S} {\bibfield  {journal} {\bibinfo  {journal} {Nucl. Phys. A}\ }\textbf {\bibinfo {volume} {518}},\ \bibinfo {pages} {149} (\bibinfo {year} {1990})}\BibitemShut {NoStop}%
\bibitem [{\citenamefont {Hagner}\ \emph {et~al.}(1995)\citenamefont {Hagner}, \citenamefont {Altmann}, \citenamefont {von Feilitzsch}, \citenamefont {Oberauer}, \citenamefont {Declais},\ and\ \citenamefont {Kajfasz}}]{Hagner:1995bn}%
  \BibitemOpen
  \bibfield  {author} {\bibinfo {author} {\bibfnamefont {C.}~\bibnamefont {Hagner}}, \bibinfo {author} {\bibfnamefont {M.}~\bibnamefont {Altmann}}, \bibinfo {author} {\bibfnamefont {F.}~\bibnamefont {von Feilitzsch}}, \bibinfo {author} {\bibfnamefont {L.}~\bibnamefont {Oberauer}}, \bibinfo {author} {\bibfnamefont {Y.}~\bibnamefont {Declais}}, \ and\ \bibinfo {author} {\bibfnamefont {E.}~\bibnamefont {Kajfasz}},\ }\href {\doibase 10.1103/PhysRevD.52.1343} {\bibfield  {journal} {\bibinfo  {journal} {Phys. Rev. D}\ }\textbf {\bibinfo {volume} {52}},\ \bibinfo {pages} {1343} (\bibinfo {year} {1995})}\BibitemShut {NoStop}%
\bibitem [{\citenamefont {Oberauer}\ \emph {et~al.}(1987)\citenamefont {Oberauer}, \citenamefont {Von~Feilitzsch},\ and\ \citenamefont {Mossbauer}}]{Oberauer:1987mg}%
  \BibitemOpen
  \bibfield  {author} {\bibinfo {author} {\bibfnamefont {L.}~\bibnamefont {Oberauer}}, \bibinfo {author} {\bibfnamefont {F.}~\bibnamefont {Von~Feilitzsch}}, \ and\ \bibinfo {author} {\bibfnamefont {R.~L.}\ \bibnamefont {Mossbauer}},\ }\href {\doibase 10.1016/0370-2693(87)90169-9} {\bibfield  {journal} {\bibinfo  {journal} {Phys. Lett. B}\ }\textbf {\bibinfo {volume} {198}},\ \bibinfo {pages} {113} (\bibinfo {year} {1987})}\BibitemShut {NoStop}%
\bibitem [{\citenamefont {Kopeikin}\ \emph {et~al.}(1990)\citenamefont {Kopeikin}, \citenamefont {Mikaelyan},\ and\ \citenamefont {Fayans}}]{Kopeikin:1990vp}%
  \BibitemOpen
  \bibfield  {author} {\bibinfo {author} {\bibfnamefont {V.~I.}\ \bibnamefont {Kopeikin}}, \bibinfo {author} {\bibfnamefont {L.~A.}\ \bibnamefont {Mikaelyan}}, \ and\ \bibinfo {author} {\bibfnamefont {S.~A.}\ \bibnamefont {Fayans}},\ }\href@noop {} {\bibfield  {journal} {\bibinfo  {journal} {JETP Lett.}\ }\textbf {\bibinfo {volume} {51}},\ \bibinfo {pages} {86} (\bibinfo {year} {1990})}\BibitemShut {NoStop}%
\bibitem [{\citenamefont {Bryman}\ and\ \citenamefont {Shrock}(2019)}]{Bryman:2019bjg}%
  \BibitemOpen
  \bibfield  {author} {\bibinfo {author} {\bibfnamefont {D.~A.}\ \bibnamefont {Bryman}}\ and\ \bibinfo {author} {\bibfnamefont {R.}~\bibnamefont {Shrock}},\ }\href {\doibase 10.1103/PhysRevD.100.073011} {\bibfield  {journal} {\bibinfo  {journal} {Phys. Rev. D}\ }\textbf {\bibinfo {volume} {100}},\ \bibinfo {pages} {073011} (\bibinfo {year} {2019})},\ \Eprint {http://arxiv.org/abs/1909.11198} {arXiv:1909.11198 [hep-ph]} \BibitemShut {NoStop}%
\bibitem [{\citenamefont {de~Gouv\^ea}\ and\ \citenamefont {Kobach}(2016)}]{deGouvea:2015euy}%
  \BibitemOpen
  \bibfield  {author} {\bibinfo {author} {\bibfnamefont {A.}~\bibnamefont {de~Gouv\^ea}}\ and\ \bibinfo {author} {\bibfnamefont {A.}~\bibnamefont {Kobach}},\ }\href {\doibase 10.1103/PhysRevD.93.033005} {\bibfield  {journal} {\bibinfo  {journal} {Phys. Rev. D}\ }\textbf {\bibinfo {volume} {93}},\ \bibinfo {pages} {033005} (\bibinfo {year} {2016})},\ \Eprint {http://arxiv.org/abs/1511.00683} {arXiv:1511.00683 [hep-ph]} \BibitemShut {NoStop}%
\bibitem [{\citenamefont {Grossheim}\ \emph {et~al.}(2009)\citenamefont {Grossheim} \emph {et~al.}}]{TWIST:2009zko}%
  \BibitemOpen
  \bibfield  {author} {\bibinfo {author} {\bibfnamefont {A.}~\bibnamefont {Grossheim}} \emph {et~al.} (\bibinfo {collaboration} {TWIST}),\ }\href {\doibase 10.1103/PhysRevD.80.052012} {\bibfield  {journal} {\bibinfo  {journal} {Phys. Rev. D}\ }\textbf {\bibinfo {volume} {80}},\ \bibinfo {pages} {052012} (\bibinfo {year} {2009})},\ \Eprint {http://arxiv.org/abs/0908.4270} {arXiv:0908.4270 [hep-ex]} \BibitemShut {NoStop}%
\bibitem [{\citenamefont {Bayes}(2013)}]{Bayes:2013esa}%
  \BibitemOpen
  \bibfield  {author} {\bibinfo {author} {\bibfnamefont {R.}~\bibnamefont {Bayes}} (\bibinfo {collaboration} {TWIST}),\ }\href {\doibase 10.1088/1742-6596/408/1/012071} {\bibfield  {journal} {\bibinfo  {journal} {J. Phys. Conf. Ser.}\ }\textbf {\bibinfo {volume} {408}},\ \bibinfo {pages} {012071} (\bibinfo {year} {2013})}\BibitemShut {NoStop}%
\bibitem [{\citenamefont {Drewes}\ \emph {et~al.}(2017)\citenamefont {Drewes} \emph {et~al.}}]{Drewes:2016upu}%
  \BibitemOpen
  \bibfield  {author} {\bibinfo {author} {\bibfnamefont {M.}~\bibnamefont {Drewes}} \emph {et~al.},\ }\href {\doibase 10.1088/1475-7516/2017/01/025} {\bibfield  {journal} {\bibinfo  {journal} {JCAP}\ }\textbf {\bibinfo {volume} {01}},\ \bibinfo {pages} {025} (\bibinfo {year} {2017})},\ \Eprint {http://arxiv.org/abs/1602.04816} {arXiv:1602.04816 [hep-ph]} \BibitemShut {NoStop}%
\bibitem [{\citenamefont {Boyarsky}\ \emph {et~al.}(2019)\citenamefont {Boyarsky}, \citenamefont {Drewes}, \citenamefont {Lasserre}, \citenamefont {Mertens},\ and\ \citenamefont {Ruchayskiy}}]{Boyarsky:2018tvu}%
  \BibitemOpen
  \bibfield  {author} {\bibinfo {author} {\bibfnamefont {A.}~\bibnamefont {Boyarsky}}, \bibinfo {author} {\bibfnamefont {M.}~\bibnamefont {Drewes}}, \bibinfo {author} {\bibfnamefont {T.}~\bibnamefont {Lasserre}}, \bibinfo {author} {\bibfnamefont {S.}~\bibnamefont {Mertens}}, \ and\ \bibinfo {author} {\bibfnamefont {O.}~\bibnamefont {Ruchayskiy}},\ }\href {\doibase 10.1016/j.ppnp.2018.07.004} {\bibfield  {journal} {\bibinfo  {journal} {Prog. Part. Nucl. Phys.}\ }\textbf {\bibinfo {volume} {104}},\ \bibinfo {pages} {1} (\bibinfo {year} {2019})},\ \Eprint {http://arxiv.org/abs/1807.07938} {arXiv:1807.07938 [hep-ph]} \BibitemShut {NoStop}%
\bibitem [{\citenamefont {Sabti}\ \emph {et~al.}(2020)\citenamefont {Sabti}, \citenamefont {Magalich},\ and\ \citenamefont {Filimonova}}]{Sabti:2020yrt}%
  \BibitemOpen
  \bibfield  {author} {\bibinfo {author} {\bibfnamefont {N.}~\bibnamefont {Sabti}}, \bibinfo {author} {\bibfnamefont {A.}~\bibnamefont {Magalich}}, \ and\ \bibinfo {author} {\bibfnamefont {A.}~\bibnamefont {Filimonova}},\ }\href {\doibase 10.1088/1475-7516/2020/11/056} {\bibfield  {journal} {\bibinfo  {journal} {JCAP}\ }\textbf {\bibinfo {volume} {11}},\ \bibinfo {pages} {056} (\bibinfo {year} {2020})},\ \Eprint {http://arxiv.org/abs/2006.07387} {arXiv:2006.07387 [hep-ph]} \BibitemShut {NoStop}%
\bibitem [{\citenamefont {Baranov}\ \emph {et~al.}(1993)\citenamefont {Baranov} \emph {et~al.}}]{Baranov:1992vq}%
  \BibitemOpen
  \bibfield  {author} {\bibinfo {author} {\bibfnamefont {S.~A.}\ \bibnamefont {Baranov}} \emph {et~al.},\ }\href {\doibase 10.1016/0370-2693(93)90405-7} {\bibfield  {journal} {\bibinfo  {journal} {Phys. Lett.}\ }\textbf {\bibinfo {volume} {B302}},\ \bibinfo {pages} {336} (\bibinfo {year} {1993})}\BibitemShut {NoStop}%
\bibitem [{\citenamefont {Bernardi}\ \emph {et~al.}(1988)\citenamefont {Bernardi} \emph {et~al.}}]{Bernardi:1987ek}%
  \BibitemOpen
  \bibfield  {author} {\bibinfo {author} {\bibfnamefont {G.}~\bibnamefont {Bernardi}} \emph {et~al.},\ }\href {\doibase 10.1016/0370-2693(88)90563-1} {\bibfield  {journal} {\bibinfo  {journal} {Phys. Lett. B}\ }\textbf {\bibinfo {volume} {203}},\ \bibinfo {pages} {332} (\bibinfo {year} {1988})}\BibitemShut {NoStop}%
\bibitem [{\citenamefont {Calaprice}\ and\ \citenamefont {Millener}(1983)}]{Calaprice:1983qn}%
  \BibitemOpen
  \bibfield  {author} {\bibinfo {author} {\bibfnamefont {F.~P.}\ \bibnamefont {Calaprice}}\ and\ \bibinfo {author} {\bibfnamefont {D.~J.}\ \bibnamefont {Millener}},\ }\href {\doibase 10.1103/PhysRevC.27.1175} {\bibfield  {journal} {\bibinfo  {journal} {Phys. Rev. C}\ }\textbf {\bibinfo {volume} {27}},\ \bibinfo {pages} {1175} (\bibinfo {year} {1983})}\BibitemShut {NoStop}%
\bibitem [{\citenamefont {Derbin}\ \emph {et~al.}(1993)\citenamefont {Derbin}, \citenamefont {Chernyi}, \citenamefont {Popeko}, \citenamefont {Muratova}, \citenamefont {Shishkina},\ and\ \citenamefont {Bakhlanov}}]{Derbin:1993wy}%
  \BibitemOpen
  \bibfield  {author} {\bibinfo {author} {\bibfnamefont {A.~I.}\ \bibnamefont {Derbin}}, \bibinfo {author} {\bibfnamefont {A.~V.}\ \bibnamefont {Chernyi}}, \bibinfo {author} {\bibfnamefont {L.~A.}\ \bibnamefont {Popeko}}, \bibinfo {author} {\bibfnamefont {V.~N.}\ \bibnamefont {Muratova}}, \bibinfo {author} {\bibfnamefont {G.~A.}\ \bibnamefont {Shishkina}}, \ and\ \bibinfo {author} {\bibfnamefont {S.~I.}\ \bibnamefont {Bakhlanov}},\ }\href@noop {} {\bibfield  {journal} {\bibinfo  {journal} {JETP Lett.}\ }\textbf {\bibinfo {volume} {57}},\ \bibinfo {pages} {768} (\bibinfo {year} {1993})}\BibitemShut {NoStop}%
\bibitem [{\citenamefont {Aguilar-Arevalo}\ \emph {et~al.}(2018)\citenamefont {Aguilar-Arevalo} \emph {et~al.}}]{PIENU:2017wbj}%
  \BibitemOpen
  \bibfield  {author} {\bibinfo {author} {\bibfnamefont {A.}~\bibnamefont {Aguilar-Arevalo}} \emph {et~al.} (\bibinfo {collaboration} {PIENU}),\ }\href {\doibase 10.1103/PhysRevD.97.072012} {\bibfield  {journal} {\bibinfo  {journal} {Phys. Rev. D}\ }\textbf {\bibinfo {volume} {97}},\ \bibinfo {pages} {072012} (\bibinfo {year} {2018})},\ \Eprint {http://arxiv.org/abs/1712.03275} {arXiv:1712.03275 [hep-ex]} \BibitemShut {NoStop}%
\bibitem [{\citenamefont {Bolton}\ \emph {et~al.}(2020)\citenamefont {Bolton}, \citenamefont {Deppisch},\ and\ \citenamefont {Bhupal~Dev}}]{Bolton:2019pcu}%
  \BibitemOpen
  \bibfield  {author} {\bibinfo {author} {\bibfnamefont {P.~D.}\ \bibnamefont {Bolton}}, \bibinfo {author} {\bibfnamefont {F.~F.}\ \bibnamefont {Deppisch}}, \ and\ \bibinfo {author} {\bibfnamefont {P.~S.}\ \bibnamefont {Bhupal~Dev}},\ }\href {\doibase 10.1007/JHEP03(2020)170} {\bibfield  {journal} {\bibinfo  {journal} {JHEP}\ }\textbf {\bibinfo {volume} {03}},\ \bibinfo {pages} {170} (\bibinfo {year} {2020})},\ \Eprint {http://arxiv.org/abs/1912.03058} {arXiv:1912.03058 [hep-ph]} \BibitemShut {NoStop}%
\bibitem [{\citenamefont {Apilluelo}\ \emph {et~al.}(2025{\natexlab{a}})\citenamefont {Apilluelo} \emph {et~al.}}]{LiquidO:2024piw}%
  \BibitemOpen
  \bibfield  {author} {\bibinfo {author} {\bibfnamefont {J.}~\bibnamefont {Apilluelo}} \emph {et~al.} (\bibinfo {collaboration} {LiquidO}),\ }\href {\doibase 10.1016/j.nima.2024.170075} {\bibfield  {journal} {\bibinfo  {journal} {Nucl. Instrum. Meth. A}\ }\textbf {\bibinfo {volume} {1071}},\ \bibinfo {pages} {170075} (\bibinfo {year} {2025}{\natexlab{a}})},\ \Eprint {http://arxiv.org/abs/2406.13054} {arXiv:2406.13054 [physics.ins-det]} \BibitemShut {NoStop}%
\bibitem [{\citenamefont {Apilluelo}\ \emph {et~al.}(2025{\natexlab{b}})\citenamefont {Apilluelo} \emph {et~al.}}]{LiquidO:2025qia}%
  \BibitemOpen
  \bibfield  {author} {\bibinfo {author} {\bibfnamefont {J.}~\bibnamefont {Apilluelo}} \emph {et~al.} (\bibinfo {collaboration} {LiquidO}),\ }\href@noop {} {\  (\bibinfo {year} {2025}{\natexlab{b}})},\ \Eprint {http://arxiv.org/abs/2503.02541} {arXiv:2503.02541 [physics.ins-det]} \BibitemShut {NoStop}%
\end{thebibliography}%

~
\newpage
\appendix


\begin{widetext}
\section{Cross Section}

 The charged current result was previously computed by Shrock~\cite{Shrock:1981wq} and we confirm this earlier result. In the following we will present the result including the neutral current and resulting interference terms. The interaction parts of the Hamiltonian for neutral current (NC) and the charged current (CC) are
\begin{eqnarray}
    \mathcal{H}_{\mathrm{NC}}&=&\frac{G_F}{\sqrt{2}}[\overline{u}_{\nu_e}\gamma_\mu(1-\gamma_5)u_{N}][\overline{u}_e\gamma^\mu(g_V-g_A\gamma_5)v_e]\cr
    \mathcal{H}_{\mathrm{CC}}&=&\frac{G_F}{\sqrt{2}}[\overline{u}_e\gamma_\mu(1-\gamma_5)u_{N}][\overline{u}_{\nu_e}\gamma^\mu(1-\gamma_5)v_e].
\end{eqnarray}
Adding these contribution  using a Fierz transformation will give us the contribution
$$\mathcal{H}=\frac{G_F}{\sqrt{2}}[\overline{u}_{\nu_e}\gamma^\mu(1-\gamma_5)u_{N}]\{\overline{u}_e\gamma_\mu[(g_V+1)-(g_A+1)\gamma_5]v_e\}.$$
 
\noindent To express the transition probability, we write
\begin{equation}
    d\Gamma=\frac{d P}{T}=\frac{d^3\mathbf{l}}{2l^0(2\pi)^3}\int\frac{d^3\bk}{2k^0(2\pi)^3}\int\frac{d^3\bq}{2q^0(2\pi)^3}\frac{G_F^2(2\pi)^4}{2}\delta^4(q+k+l-p)\frac{\sum_{\mathrm{spin}}|M|^2}{2E_{N}},
\end{equation}
where $q$, $k$, $l$, and $p$ are the four-momentum of the outgoing electron, positron, neutrino and the incoming heavy neutrino correspondingly. The $\frac{1}{2}$ in the front arises from averaging over the two possible spin-states of $N$. The squared invariant matrix element $M$ is defined to be
\begin{equation}
    M=[\overline{u}_{\nu_e}\gamma^\mu(1-\gamma_5)u_{N}e^{i(l-p)x}]\{\overline{u}_e\gamma_\mu[(g_V+1)-(g_A+1)\gamma_5]v_ee^{i(q+k)x}\}.
\end{equation}
To evaluate the transition rate $d\Gamma$, we need to first evaluate $\sum_{\mathrm{spin}}|M|^2=\sum_{\mathrm{spin}}M^\dagger M$, which can be done in two parts separately for neutrinos and charged leptons, respectively. Focusing on the neutrino part first,
\begin{eqnarray}
    &&\sum_{t}[\overline{u}_{\nu_e}(\Vec{l},t)\gamma^\mu(1-\gamma_5)u_{N}(\Vec{p},s)][\overline{u}_{\nu_e}(\Vec{l},t)\gamma^\nu (1-\gamma_5)u_{N}(\Vec{p},s)]^\dagger\cr
    &=&\sum_{t}\overline{u}_{\nu_e}(\Vec{l},t)\gamma^\mu(1-\gamma_5)u_{N}(\Vec{p},s)\overline{u}_{N}(\Vec{p},s)\gamma^\nu (1-\gamma_5)u_{\nu_e}(\Vec{l},t)\cr
    &=&\Tr\bigg\{\gamma^\mu(1-\gamma_5)\lslash\gamma^\nu(1-\gamma_5)(\pslash+m_{N})\bigg(\frac{1+\gamma_5 \sslash}{2}\bigg)\bigg\}\cr
    &=&\Tr\{\gamma^\mu(1-\gamma_5)\lslash\gamma^\nu(1-\gamma_5)(\pslash+m_{N}\sslash)\}\cr
    &=&\Tr\{\lslash\gamma^\nu(\pslash+m_{N}\sslash)\gamma^\mu(1-\gamma_5)\}\cr
    &=&8[l^\mu (p+m_{N}s)^\nu -g^{\mu\nu}(l\cdot (p+m_{N}s))+l^\nu (p+m_{N}s)^\mu-i\epsilon^{\mu\alpha \nu \beta}l_\alpha (p+m_{N}s)_\beta]\cr
    &=&8[l^\mu a^\nu -g^{\mu\nu}(l\cdot a)+l^\nu a^\mu-i\epsilon^{\mu\alpha \nu \beta}l_\alpha a_\beta],
\end{eqnarray}
where $a=(p+m_{N}s)$. The final spins are summed over and the initial spin $s$ will give us insights on the angle of emission of the out-going neutrino $\nu_e$. Similarly, the charged lepton part is
\begin{eqnarray}
    &&\frac{1}{2}\sum_{t',t''}[\overline{u}_e(\Vec{q},t')\gamma_\mu\{(g_V+1)-(g_A+1)\gamma_5\}v_e(\Vec{k},t'')]\cr
    &&\qquad\qquad\qquad[\overline{u}_e(\Vec{q},t')\gamma_\mu\{(g_V+1)-(g_A+1)\gamma_5\}v_e(\Vec{k},t'')]^\dagger\cr
    &=&\sum_{t',t''}\overline{u}_e(\Vec{q},t')\gamma_\mu\{(g_V+1)-(g_A+1)\gamma_5\}v_e(\Vec{k},t'')\cr
    &&\qquad\qquad\qquad\overline{v}_e(\Vec{k},t'')\gamma_\mu\{(g_V+1)-(g_A+1)\gamma_5\}u_e(\Vec{q},t')\cr
    &=&\Tr\{\gamma_\mu[(g_V+1)-(g_A+1)\gamma_5](\qslash+m_e)\gamma_\nu[(g_V+1)-(g_A+1)\gamma_5](\kslash+m_e)\}\cr
    &=&+\Tr\{\gamma_\mu(g_V+1)(\qslash+m_e)\gamma_\nu(g_V+1)(\kslash+m_e)\}\cr
    &&-\Tr\{\gamma_\mu(g_V+1)(\qslash+m_e)\gamma_\nu(g_A+1)\gamma_5(\kslash+m_e)\}\cr
    &&-\Tr\{\gamma_\mu(g_A+1)\gamma_5(\qslash+m_e)\gamma_\nu(g_V+1)(\kslash+m_e)\}\cr
    &&+\Tr\{\gamma_\mu(g_A+1)\gamma_5(\qslash+m_e)\gamma_\nu(g_A+1)\gamma_5(\kslash+m_e)\}\cr
    &=&+\Tr\{\gamma_\mu(g_V+1)\qslash\gamma_\nu(g_V+1)\kslash\}+m_e^2\Tr\{\gamma_\mu(g_V+1)\gamma_\nu(g_V+1)\}\cr
    &&-\Tr\{\gamma_\mu(g_V+1)\qslash\gamma_\nu(g_A+1)\gamma_5\kslash\}-m_e^2\Tr\{\gamma_\mu(g_V+1)\gamma_\nu(g_A+1)\gamma_5\}\cr
    &&-\Tr\{\gamma_\mu(g_A+1)\gamma_5\qslash\gamma_\nu(g_V+1)\kslash\}-m_e^2\Tr\{\gamma_\mu(g_A+1)\gamma_5\gamma_\nu(g_V+1)\}\cr
    &&+\Tr\{\gamma_\mu(g_A+1)\gamma_5\qslash\gamma_\nu(g_A+1)\gamma_5\kslash\}+m_e^2\Tr\{\gamma_\mu(g_A+1)\gamma_5\gamma_\nu(g_A+1)\gamma_5\}\cr
    &=&+[(g_V+1)^2+(g_A+1)^2]\Tr\{\gamma_\mu\qslash\gamma_\nu\kslash\}+2(g_V+1)(g_A+1)\Tr\{\gamma_\mu\qslash\gamma_\nu\kslash\gamma_5\}\cr
    &&+m_e^2[(g_V+1)^2-(g_A+1)^2]\Tr\{\gamma_\mu\gamma_\nu\}\cr
    &=&4 X(q_\mu k_\nu-g_{\mu\nu}(q\cdot k)+q_\nu k_\mu)-8Y i\epsilon_{\mu'\alpha'\nu'\beta'}q^\alpha k^\beta+4 m_e^2 Z g_{\mu \nu},
\end{eqnarray}
where $X=[(g_V+1)^2+(g_A+1)^2]$, $Y=[(g_V+1)(g_A+1)]$, and $Z=[(g_V+1)^2-(g_A+1)^2]$. Now, putting them together we get
\begin{eqnarray}
    &&\frac{1}{2}\sum_{\mathrm{spin}}|M|^2\cr
    &=&32 \big[(X-2Y)(l\cdot q)(a\cdot k)+(X+2Y)(l\cdot k)(a\cdot q)-Z m_e^2(l\cdot a)\big].
\end{eqnarray}
To integrate over the invariant matrix element and express the result with $Q=p-l$, the momentum that goes to the $\ee$-pair, we generally need three identities. In our case, the first two are sufficient. The first identity is
\begin{eqnarray}
    I_{\alpha\beta}&=&\int \frac{d^3\mathbf{k}}{2k^0}\int\frac{d^3\mathbf{q}}{2 q^0}k_\alpha q_\beta\delta^4(k+q+l-p)\cr
    &=&\frac{\pi}{24}\sqrt{1-\frac{4m_e^2}{Q^2}}\bigg[g_{\alpha\beta} Q^2\bigg(1-\frac{4m_e^2}{Q^2}\bigg)+2Q_\alpha Q_\beta\bigg(1+\frac{2m_e^2}{Q^2}\bigg)\bigg]\Theta(Q^2-4m_e^2),
\end{eqnarray}
which can be obtained by using the ansatz $I_{\alpha\beta}=A Q^2 g_{\alpha\beta}+B Q_\alpha Q_\beta$. We will solve for $A$ and $B$  later. Further, we have
\begin{equation}
    I=\int \frac{d^3\bk}{2k^0}\int\frac{d^3 \bq}{2q^0}\delta^4(k+q+l-p)=\frac{\pi}{2}\sqrt{1-\frac{4m_e^2}{Q^2}}.
\end{equation}
In the center-of-mass system, $\bq+\bk+\bl=0$ and $q^0+k^0+l^0=m_\nuh$ hold and for the neutrino we also have $l^0=|\bl|$, further we define $\bl\cdot\bk=l^0 k^0\cos\alpha$. Then, the third identity is
\begin{eqnarray}
    I'&=&\int \frac{d^3\bk}{2k^0}\int\frac{d^3 \bq}{2q^0}\delta^4(k+q+l-p)(l\cdot k)\cr
    &=&\int \frac{d^3\bk}{2k^0}\frac{\delta(\sqrt{|\bk+\bl|^2+m_e^2}+k^0+l^0-m_\nuh)}{\sqrt{|\bk+\bl|^2+m_e^2}}(l^0 k^0)(1-\cos\alpha)\cr
    &=&l^0\int d\Omega(1-\cos\alpha) \int\frac{(k^0)^2 d k^0}{2}\frac{\delta(\sqrt{(k^{0})^2+|\bl|^2+2 k^0|\bl|+m_e^2}+k^0+l^0-m_\nuh)}{\sqrt{(k^{0})^2+|\bl|^2+2 k^0|\bl|+m_e^2}}\cr
    &=&4\pi l^0\bigg(\frac{(k^0)^2}{2\sqrt{(k^{0})^2+|\bl|^2+2 k^0|\bl|+m_e^2}(1+\frac{k^0+|\bl|}{\sqrt{(k^0+|\bl|)^2+m_e^2}})}\bigg)\bigg|_{k^0=\frac{-m_e^2-2|\bl|m_\nuh+m_\nuh^2}{2m_\nuh}}\cr
    &=&4\pi l^0\frac{(m_e^2-Q^2)^2}{8m_\nuh^3}\cr
    &=&\frac{\pi}{4m_\nuh^4}(m_\nuh^2-Q^2)(m_e^2-Q^2)^2.
\end{eqnarray}
Here, the $\delta^4$ was split up into a spatial part and a time part so we can integrate over them  by using $d^3\bk=|\bk|^2d\Omega\frac{k_0}{|\bk|}dk_0$. The spin for the incoming heavy neutrino in the center-of-mass frame is simply $(\frac{\Vec{s}\cdot\Vec{p}}{m_{N}},\Vec{s}+\frac{(\Vec{p}\cdot \Vec{s})\Vec{p}}{2m_{N}}^2)=(0,\Vec{s})$. For these identities, the trick is to split up the $\delta^4$ like mentioned and to directly integrate over the spatial coordinates. Since the interaction has to be time-like we need to integrate considering only the frame where $Q$ has only the time component $\Tilde{Q}^0$. Once integrated, we can transform back into the original frame. Exploiting Lorentz-invariance and using the first identity we obtain
$$Q^2=m_\nuh^2-2m_\nuh l^0\Rightarrow l^0=\frac{1}{2m_\nuh}(m_\nuh^2-Q^2)$$
$$(a\cdot l)=m_{N}l^0+m_\nuh |\Vec{s}|\cos\theta|\Vec{l}|=\frac{1}{2}(m_\nuh^2-Q^2)+\frac{1}{2}|\Vec{s}|\cos\theta(m_\nuh^2-Q^2)$$
$$Q\cdot l=m_{N}l^0=p\cdot l=\frac{1}{2}(m_\nuh^2-Q^2) $$
\begin{equation*}
    \begin{split}
        Q\cdot a=Q\cdot p+m_\nuh Q\cdot s&=m_\nuh^2-m_\nuh l^0-m_\nuh|\Vec{s}|\cos\theta|\Vec{l}|\\
        &=m_\nuh^2-\frac{1}{2}(m_\nuh^2-Q^2)-\frac{1}{2}|\Vec{s}|\cos\theta(m_\nuh^2-Q^2).
    \end{split}
\end{equation*}

\bigskip
\noindent
Once the other identities are applied, we can see that $(l\cdot q)[(p+m_{N}s)\cdot k]$ has been integrated to become 
\begin{eqnarray}
    &&\frac{\pi}{24}\sqrt{1-\frac{4m_e^2}{Q^2}}\bigg[l\cdot (p+m_\nuh s) Q^2\bigg(1-\frac{4m_e^2}{Q^2}\bigg)+2(Q\cdot l) (Q\cdot (p+m_\nuh s))\bigg(1+\frac{2m_e^2}{Q^2}\bigg)\bigg]\cr
    &=&\frac{\pi}{48}(1-Q^2)\frac{1}{Q^2}\sqrt{1-\frac{4m_e^2}{Q^2}}\bigg[Q^2+2Q^4-2m_e^2(Q^2-1)\cr
    &&\qquad\qquad\qquad\qquad\qquad\qquad-|\Vec{s}|\cos\theta[Q^2-2Q^4+2m^2(1+Q^2)]\bigg]
\end{eqnarray}
And $(l\cdot k)(p+m_\nuh s)\cdot q$ is integrated to be 
\begin{eqnarray}
    &&\frac{\pi}{48}(1-Q^2)\frac{1}{Q^2}\sqrt{1-\frac{4m_e^2}{Q^2}}\bigg[Q^2+2Q^4-2m_e^2(Q^2-1)\cr
    &&\qquad\qquad\qquad\qquad\qquad\qquad-|\Vec{s}|\cos\theta[Q^2-2Q^4+2m^2(1+Q^2)]\bigg].
\end{eqnarray}
Same as above. Lastly, the third term
\begin{eqnarray}
    &&\frac{\pi}{2}\sqrt{1-\frac{4m_e^2}{Q^2}}m_e^2(l\cdot[p+m_{\nuh}s])\cr
    &=&\frac{\pi}{2}\sqrt{1-\frac{4m_e^2}{Q^2}}m_e^2(m_\nuh l^0+m_\nuh |\Vec{s}|\cos\theta l^0)\cr
    &=&\frac{\pi}{4}\sqrt{1-\frac{4m_e^2}{Q^2}}m_e^2(m_\nuh^2-Q^2)(1+|\Vec{s}|\cos\theta),
\end{eqnarray}
where $l^0=|\Vec{l}|$ and $\Vec{s}\cdot \Vec{l}=|\Vec{s}|\cos\theta|\Vec{l}|$ that defines the angle of emission of $\nu_e$ relative to $N$-polarization direction $\Vec{s}$ with  furthermore, $p\cdot l=m_{N}l^0$.
Now we put everything back together and write
\begin{eqnarray}
    \frac{d\Gamma}{ d\cos\theta}
    &=&\frac{2\pi \bl^2 d l^0}{2 l^0(2\pi)^3}\frac{G_F^2(2\pi)^4}{(2\pi)^6}\frac{1}{2 m_{N}}\int\frac{d^3\bk}{2k^0}\int\frac{d^3\bq}{2q^0}\delta^4(q+k+l-p)\frac{1}{2}\sum_{\mathrm{spin}}|M|^2\cr
    \frac{d^2\Gamma}{dl^0 d\cos\theta}&=&\frac{G_F^2 m_{N}^5}{192\pi^3}(1-Q^2)^2\sqrt{1-\frac{4m_e^2}{Q^2}}\frac{1}{Q^2}\bigg\{\cr
    &&\qquad\qquad\bigg[2X\bigg(Q^2+2Q^4-2m_e^2(Q^2-1)\bigg)-12Z Q^2 m_e^2\bigg]\cr
    &&-|\Vec{s}|\cos\theta\bigg[2X\bigg(Q^2-2Q^4+2m^2(1+Q^2)\bigg)+12Z Q^2 m_e^2\bigg]\bigg\},
\end{eqnarray}
where, as a reminder, $X=[(g_V+1)^2+(g_A+1)^2]$ and $Z=[(g_V+1)^2-(g_A+1)^2]$. Finally, we will change to the format used by Shrock~\cite{Shrock:1981wq} 
with the following definitions
$$\Gamma_0=\frac{G_F^2 m_{N}^5}{192\pi^3},\qquad \frac{d^2\Gamma}{d l^0 d\cos\theta}=\Gamma_0|U_{s1}|^2\frac{d^2\overline{\Gamma}}{d l^0 d\cos\theta}$$
\begin{eqnarray}
    \frac{d^2\overline{\Gamma}}{d l^0 d\cos\theta}&=&2(1-Q^2)^2\sqrt{1-\frac{4m_e^2}{Q^2}}\frac{1}{Q^2}\bigg\{\cr
    &&\qquad\qquad\bigg[X\bigg(Q^2+2Q^4-2m_e^2(Q^2-1)\bigg)-6Z Q^2 m_e^2\bigg]\cr
    &&-|\Vec{s}|\cos\theta\bigg[X\bigg(Q^2-2Q^4+2m_e^2(1+Q^2)\bigg)+6Z Q^2 m_e^2\bigg]\bigg\}.
\end{eqnarray}

Setting up the decay rate, we integrate the momentum over the initial and final states. The interaction part of the Lagrangian constitutes four massive dimensions, so overall we have mass dimension $5$. We then pull out $m_{N}^5$ such that every massive term will be in the dimension of the heavy neutrino. On the other hand, since we are interested in the polarization of the initial state, we only sum over the spins of the final states. 

\section{Electron and Positron Opening Angle}
\label{app:angle}

We can generally write the kinematics of the three body decay $\nu_H\rightarrow e^+ e^- \nu_e$ in four-momentum, independent of which channel the decay happens through. Then, writing in Lorentz-invariant form, we will be able to set up the kinematics equation that can be solved to obtain the opening angle of the $e^-e^+$ pair. To begin with, we denote the four-momentum of the heavy neutral lepton $\nu_H$, outgoing neutrino $\nu_e$, electron $e^-$ and positron $e^+$ to be $p$, $l$, $q$ and $k$, respectively. The squared of the momentum transfer to $e^-e^+$ is
$$(p-l)^2=(q+k)^2.$$
Now, we rewrite everything in the lab frame where observations are made
$$m_H^2-2(E_H E_\nu-p_H p_\nu \cos\theta)=2m_e^2+2(E_{e^+}E_{e^-}-p_{e^+}p_{e^-}\cos\alpha).$$
Here, we assumed that the outgoing neutrino mass is zero. The opening angle of $e^+ e^-$ is $\alpha$ and the angle between $\nu_e$ and $\nu_H$ is $\theta$.
Solving the equation for the $\cos\alpha$
$$\cos\alpha=\frac{2me^2+E_{e^+}E_{e^-}-m_{\nu_H}^2-2E_H E_\nu-2\sqrt{E_H^2-m_H^2}E_\nu\cos\theta}{p_{e^+}p_{e^-}},$$
where $E_\nu=E_H-E_{e^+}-E_{e^-}$ and $p_{e^\pm}=\sqrt{E_{e^\pm}^2-me^2}$. For the  parameter space of interest delineated by the electron mass and the upper energy of the neutrino flux from the Sun, we have $2\,\mathrm{MeV}\leq E_H\leq 16$\,MeV, which indicates that $E_{e^+}+E_{e^-}+E_\nu\leq 16\,\mathrm{MeV}$ sets an upper limit on the  outgoing energies $E_{e^+}$ and $E_{e^-}$. On the other hand, $\cos\theta$ will be in the  range $\cos\theta_{\mathrm{min}}\leq\cos\theta\leq 1$, where
$$\cos\theta_{\mathrm{min}}=\frac{1}{\sqrt{1-(\frac{m_H}{E_H})^2}}\Bigg(\frac{m_H}{\frac{m_H^2-4m_e^2}{2m_H}}\bigg(1-\frac{E_{e^+}+E_{e^-}}{E_H}\bigg)-1\Bigg).$$
Now, if we fix the four-momentum of the heavy neutral lepton, the opening angle will depend on the energies of the $e^+$ and $e^-$. It is natural to expect that when the Lorentz-boost is larger, the possible minimum $e^+e^-$'s opening angle will be smaller, depending on the direction of boosting. With careful study of opening angle for different momenta of $\nu_H$ and $\nu_e$, we found that the bigger the three momentum $\Vec{p}_H$ is, the smaller $\alpha$ will be. Note that $\cos\theta$ does not change the possible smallest $\alpha$, it just changes the probability for the occurrence of each $\alpha$. For example, if the smallest possible $\alpha=30^\circ$, then, a smaller $\cos\theta$ results in a higher probability  to observe $\alpha=30^\circ$. The smallest possible $\alpha$ we found by varying all the parameters is approximately $\alpha\gtrsim 50^\circ$. Therefore, there  will be a clear $e^+ e^-$ signature with two well-separated tracks.

\end{widetext}


\end{document}